\documentclass[12pt]{iopart}

\usepackage[hidelinks,colorlinks=true,linkcolor=blue,urlcolor=blue,citecolor=blue,anchorcolor=blue]{hyperref}
\usepackage{graphicx}
\usepackage[mathscr]{euscript}
\usepackage{cite}
\usepackage{multirow}
\begin{document}

\title[]{Can a vector beam be critically coupled leading to perfect absorption?
}

\author{Sauvik Roy$^{1,\footnotemark[1]}$, Nirmalya Ghosh$^1$, Ayan Banerjee$^1$, and Subhasish Dutta Gupta$^{1,2}$}
\address{$^1$Department of Physical Sciences, IISER-Kolkata, Mohanpur 741246, India. \\  $^2$Tata Institute of Fundamental Research, Hyderabad, Telangana 500046, India. }

\ead{\mailto{sauvikroy3388@gmail.com} $^{\footnotemark[1]}$, \mailto{ayan@iiserkol.ac.in}, and \mailto{sdghyderabad@gmail.com}.}

\vspace{10pt}

\begin{abstract}
Critical coupling has emerged as a prominent area of research in recent years. However, most theoretical models are based on scalar theories (and occasionally coupled mode theories), which inadequately account for the polarization states of the incident light. To bridge this gap, we revisit the concept of critical coupling in planar multilayer structures using a full vectorial theory, where conventional plane wave illumination is replaced by well-defined vector beams with and without orbital angular momentum (OAM). Our investigation explores the possibility of complete absorption of monochromatic beams without and with intrinsic OAM  (such as Gaussian and Laguerre-Gaussian (LG)),  incident on the multilayer structure at normal or oblique incidence. A two-component metal-dielectric composite film is chosen as the absorbing layer in the system. Our results demonstrate a significant reduction in the intensities of the reflected and transmitted beams at normal incidence, with reduced efficiency for oblique incidence due to the lack of spatial overlap of multiply reflected components. Interestingly, we also observe super-scattering from the same structures when conditions for constructive interference of the various reflected components are satisfied. This work highlights the need to incorporate the vector nature of beams by retaining the complete polarization information of off-axis spatial harmonics in future studies.
\end{abstract}

\footnotetext[1]{Corresponding author}

\section{Introduction}

Critical coupling \cite{TISCHLER200794, Tischler06, DuttaGupta07, Deb2007, doi:10.1126sciadv.1701377, Guo:21, xiang2014critical, Li:17, wgmmodesilicamicrosphere,blackabsorbervisible} is a mechanism that enables optimal energy transfer between two coupled resonant components in a complex system. In optics, critical coupling results in simultaneous null transmission and reflection from a material (typically resonators with finite losses) specifically designed for near-total absorption of the incoming electromagnetic energy. In this context, there have been studies \cite{TISCHLER200794,Tischler06} of heterolayers with a $5nm$ thin lossy J-aggregated film with high oscillator strength reporting critical coupling ($\approx97\%$ absorption) of the incident field at a wavelength $\lambda = 591 nm$. Dutta Gupta et al. leveraged the frequency tunability of a metal-dielectric composite thin film replacing the polymer absorbing layer \cite{DuttaGupta07}, and later extended the notion of critical coupling to the case of oblique incidence \cite{Deb2007} for both TE and TM polarizations. Several other structures involving negative index metamaterials \cite {duttagupta2010oc,DuttaGupta2010joo}, graphene-based chiral metamaterials  \cite{doi:10.1126sciadv.1701377,Guo:21}, hyperbolic metamaterials \cite{xiang2014critical}, monolayer transition-metal dichalcogenides \cite{Li:17}, whispering-gallery modes (WGM) in silica microspheres coupled with a tapered fiber \cite{wgmmodesilicamicrosphere}, lossy quantum systems \cite{lossyquantumsystems}, etc. also achieved critical coupling.
	\par
Note that the extension of CC (with single channel input) to dual channel prototype is the well-known coherent perfect absorption (CPA) (also referred to as antilaser), which has been studied extensively  \cite{wancpa2011,timereversedlaser,DuttaGupta:12,baranov2017coherent,longhi2011, Jeffers2000, barnett1998, agarwal2014} in the past two decades. It is worth mentioning that significant progress has been achieved in extending the concept of (single-channel) critical coupling or (double-channel) coherent perfect absorption (CPA) to arbitrary superpositions of plane waves in both single and multichannel input scenarios \cite{arbitrary,ep,esurface,cpaep,arbwave}. These advancements harness the principles of exceptional points (EPs) and exceptional surfaces (ESs), often relying on highly degenerate cavities that demand meticulous tuning of specially designed non-Hermitian coupled cavity systems, either with or without intracavity lenses. Nevertheless, these studies predominantly employed a scalar coupled mode theory, overlooking a comprehensive treatment of the polarization states of the incident light. Similar shortcomings have recently been identified and successfully addressed in the context of coherent perfect absorption (CPA) where two structured beams illuminated a metal-dielectric composite film from opposite sides \cite{olcpabeam,DuttaGupta:12}.
 \par
To the best of our knowledge, critical coupling \cite{DuttaGupta07, Deb2007,PhysRevApplied.3.037001, TISCHLER200794, Tischler06}, till date, incorporates single plane wave description without adequate input about its polarization state both for the normal and oblique incidence cases. A simple example of input circularly polarized light may illustrate the difficulty. Thus, for oblique incidence, the CC for TE and TM occurs at different frequencies for the same structure, thus making it nearly impossible to meet the conditions of CC for both the polarizations simultaneously, leading to issues in the case of circularly polarized light.  The case of input vector beams having finite transverse extents and carrying orbital angular momentum is even more complicated because of the off-axis spatial harmonic components. In general, beams with a finite cross-section can only be represented as a collection of plane waves. While interacting with planar interfaces, these plane waves not only make different `angles of incidence' but also belong to different `planes of incidence'. Moreover, the polarizations of these plane waves in the beam spectrum are different from each other \cite{Bliokh2013}. In short, this explains -- at least qualitatively -- why perfect critical coupling cannot be achieved for a beam in structures with planar geometry. More broadly, the critical coupling mechanism has limitations in real applications due to the combination of the limited overlap (of the reflected beam components from different interfaces within the planar structure) imposed by the field’s vectorial nature and the specific geometry of the incident beam. Given the absence of a full vectorial theory, we revisit critical coupling where the incident plane waves are now substituted with well-defined vector beams, either with or without OAM. Importantly, incorporating polarization in the incoming beams, especially those with orbital angular momentum (OAM), adds a new dimension that enriches critical coupling through spin-orbit interaction effects, such as the spin-Hall effect and orbital-Hall effect, leading to large in-plane and out-of-plane shifts.
\par
To highlight the vulnerability of the coupling mechanism for vector beams, we have chosen the widely studied multilayer structure used for critical coupling, which incorporates a top absorbing thin layer separated from a distributed Bragg reflector (DBR) by a spacer layer. A metal-dielectric (silver-silica) composite layer is chosen as the absorbing layer. The ability to tune the resonance position and absorption strength in a metal-dielectric composite, simply by varying the metal volume fraction, is particularly leveraged to adjust the critical coupling frequency in the structure. It is to be noted that the choice of the structure is motivated by several earlier studies \cite{DuttaGupta07,Deb2007} where the same structure has been reported to exhibit critical coupling under both normal and oblique incidence of a single plane wave. With this structure, we have studied the characteristics of the reflected and transmitted beams when illuminated by polarized Gaussian and LG beams -- both for the normal and oblique incidences under critical coupling conditions for the central wave vector. It has been revealed that although perfect critical coupling cannot be achieved for beams, significant attenuation of the reflected and transmission beam intensities is attainable for normal incidence. The residual reflected and transmitted beams exhibit exotic structures governed by the polarization and OAM content of the incident beam. We also numerically evaluate the in-plane Goos-Hänchen (GH), out-of-plane Imbert-Fedorov (IF), and vortex-induced shifts of these beams.
 
\section{Theoretical formalism}

\begin{figure}[ht!]
\centering\includegraphics[width=0.75 \linewidth]{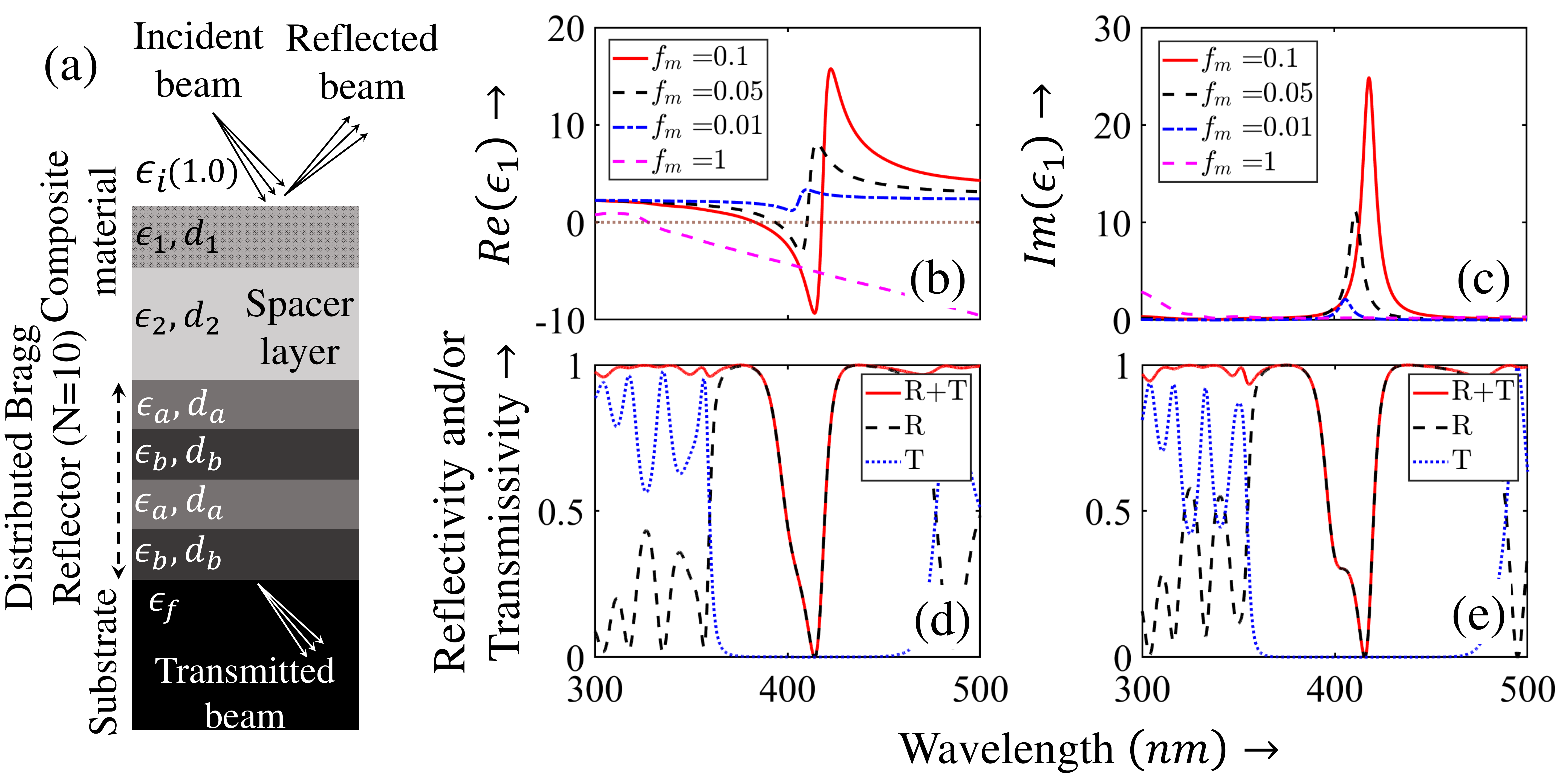}
\caption{(a) Schematic of the multilayered medium illuminated by a vector beam. (b) Real and (c) imaginary parts of the permittivity ($\epsilon_1$) of the composite material for various volume fractions ($\epsilon_m$) of silver. Total scattering R + T (red solid lines) as a function of wavelength for (d) normally and (e) obliquely incident ($\theta=45^{\circ}$) $s$-polarized plane waves.}
\label{response}
\end{figure}

Our structure is a four-component (Fig. \ref{response} (a)) optical system comprising of 1) a metal-dielectric composite layer of dielectric constant $\epsilon_1$ and thickness $d_1$ , 2) a spacer layer of dielectric constant $\epsilon_2$ and thickness $d_2$, 3) a 2N+1 layered distributed Bragg reflector (DBR), with alternating dielectric constant $\epsilon_a$, $\epsilon_b$ and thickness $d_a$ and $d_b$, respectively, and 4) a silica substrate (considered to be semi-infinite with dielectric constant $\epsilon_f$). The essential requirement of critical coupling is that the absorption dip of the composite material must lie within the stopgap of the DBR \cite{DuttaGupta07,Tischler06,TISCHLER200794,Deb2007}. In structures with fixed DBR parameters (such as widths and dielectric functions), the band-gap shifts to lower wavelengths as the angle of incidence increases \cite{Deb2007}. In contrast, the absorption dip of the composite material remains unaffected. Thus, for oblique incidence \cite{Deb2007}, the absorption frequency may fall outside the band-gap, leading to loss of critical coupling. To avoid this, it is imperative to adjust the thicknesses ($d_a$ and $d_b$) of the DBR sublayers such that each sublayer corresponds to a $\lambda/4$ plate at each angle of incidence $\theta$, thus always maintaining a central Bragg frequency fixed, say, at $\lambda_c = 410nm$. The thicknesses of the DBR sublayers are given as:
\begin{equation} \label{DBR_widths}
d_{a,b}=\frac{\lambda_c}{4n_{a,b}\cos{\theta_{a,b}}},\qquad  \theta_{a,b}=\sin^{-1}{\left[\frac{n_i}{n_{a,b}}\sin{\theta}\right]} \\
\end{equation}
where, $n_i=\sqrt{\epsilon_i}$ and $n_{a,b}=\sqrt{\epsilon_{a,b}}$. In addition to the DBR, the other crucial component is the metal-dielectric composite film. For this, we have considered silver nanoparticles (dielectric constant $\varepsilon_m$) embedded in a silica host (dielectric constant $\varepsilon_h$). It has been shown earlier that the Maxwell-Garnett (MG) theory \cite{Garnett} can be applied effectively to describe the response (dispersion and absorption) in the visible range with particle sizes not exceeding $10nm$. According to this theory, the effective dielectric function of the composite medium is given by: 
\begin{equation}
\varepsilon_{1}(\omega)=\varepsilon_{h}+\frac{f_mx(\varepsilon_{m}-\varepsilon_{h})}{1+f(x-1)},\qquad   x=\frac{3\varepsilon_{h}}{\varepsilon_{m}+2\varepsilon_{h}} \\
\end{equation}
Where, $f_m$ is the volume fraction of the metal (silver). It is important to note that any abrupt variation in the field can adversely affect the quasi-static approximation of the MG theory, causing the actual dispersion to deviate from the theoretical predictions \cite{DuttaGupta07,Deb2007}. Therefore, it is crucial to exercise caution in situations where localized Mie plasmons of metal inclusions may be excited. However, in scenarios where field enhancement is moderate and particle sizes remain within acceptable limits, the MG theory remains a reliable tool for predicting the effective dielectric function. 
\begin{figure}[ht!]
\centering\includegraphics[width=0.7 \linewidth]{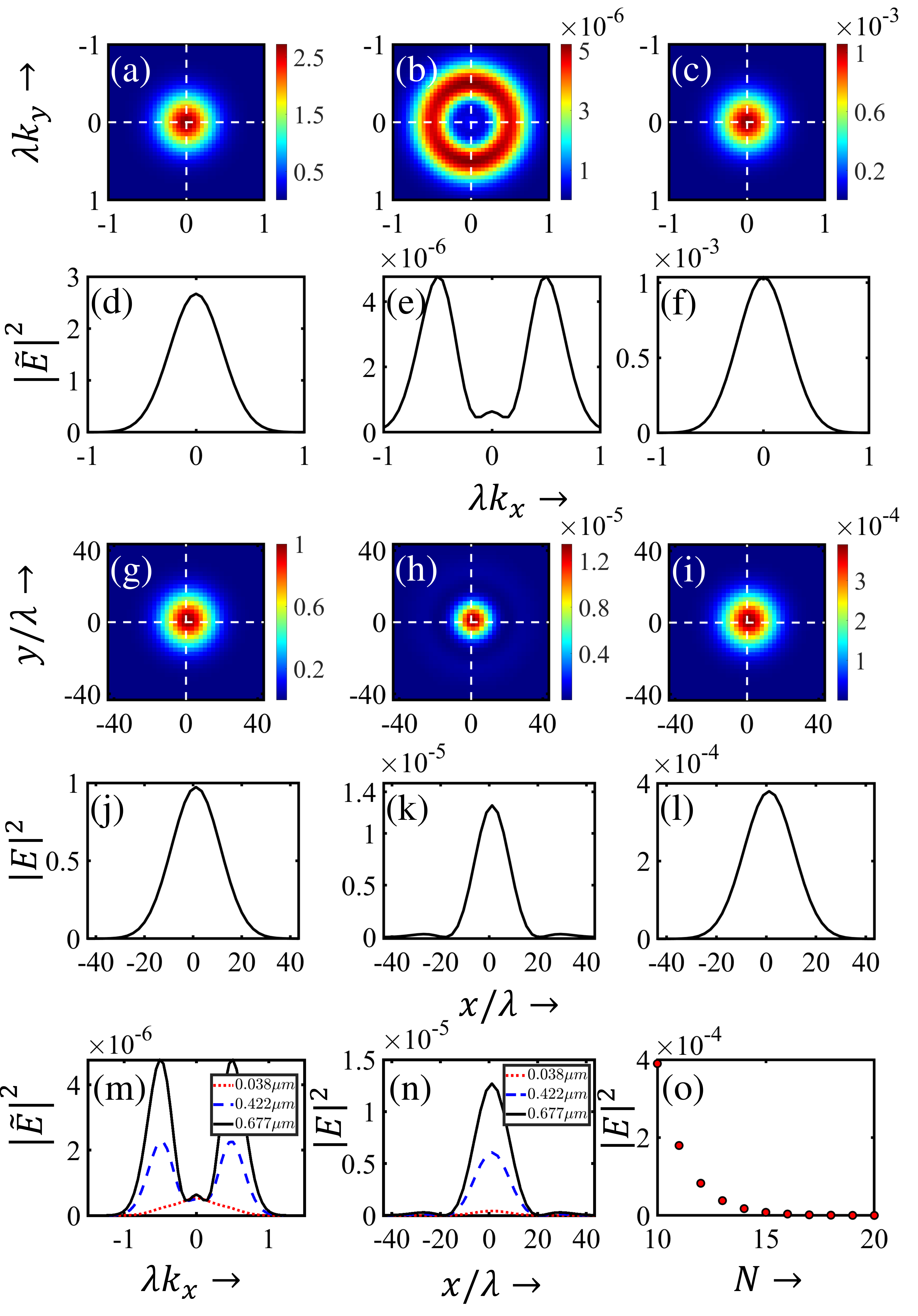}
\caption{(a) Incident, (b) reflected, and (c) transmitted spectra of the $s-$polarized Gaussian beam incident normally on the structure. The line plots in (d)-(f) along the horizontal axis further illustrate the characteristics of these spectra presented in (a)-(c). Real-space (g) incident, (h) reflected, and (i) transmitted beams. Line plots (j)-(l) along the $x-$axis depict the nature of the beams corresponding to (g)-(i). Parameters for (a)-(l) are: $\epsilon_i=1.0$, $\epsilon_2=2.6244$, $\epsilon_h=2.25$, $\epsilon_a=5.7121$, $\epsilon_b=2.6244$, $\epsilon_f=2.25$, $d_1=0.01\mu m$, $d_2=0.677\mu m$, $f_m=0.05$, N=10. (m) Reflected beam spectra and (n) the corresponding reflected beams for different spacer thicknesses: $d_2 = 0.038\mu m, 0.422\mu m, 0.677\mu m$. Other parameters are the same as mentioned earlier. (o) Maximum value of the transmitted beam with increasing period of the DBR sublayers N. }
\label{fig:gaussiannormal}
\end{figure}

Next, we invoke an angular spectrum formalism for polarized beams originally developed by Bliokh and Aillo \cite{Bliokh2013} to investigate beams transmitted through or reflected from a single interface. This method has later been adapted for use with any stratified medium \cite{SinhaBiswas2023,BISWAS2024130766}. Here, we highlight two significant drawbacks of the original formulation in \cite{Bliokh2013} and explain how they have been resolved through two substantial improvements. First, in their framework, they implemented a strict paraxial approximation, retaining only the linear variations from the central $k$-vector in the spectrum. This approximation is removed by considering a much broader spectrum retaining at least all the propagating components of a beam, extending the applicability of the formalism to practically all polarized beams beyond the strict paraxial limit. Second, the exact values of the Fresnel reflection and transmission amplitude coefficients are utilized, as opposed to the Taylor series expansion used in \cite{Bliokh2013} including only linear terms. For a multilayer (or stratified) medium, these coefficients are computed through the characteristic matrix formalism \cite{Born_Wolf,gupta2015wave,PhysRevA.105.063514}. These developments may not yield closed-form expressions for field profiles, but enhance the applicability of the method for numerical studies of tightly focused beams. Particularly, due to the inclusion of larger momentum spread, the analysis of normal incidence becomes possible as, in this case, the non-trivial contribution comes only from higher order terms.  
\par
For the incident Gaussian beam, we begin with the corresponding normalized spectrum \cite{SinhaBiswas2023,Bliokh2013}:
\begin{equation}\label{gaussain_spectrum}
    |\mathbf{E}_i\rangle = \frac{w_0}{\sqrt{2\pi}} \exp\{-(k_x^2+k_y^2)w_0^2/4\}(A_p\mathbf{e}_p + A_s\mathbf{e}_s)
\end{equation}
where, $w_0$ is the beam waist, $A_p$ and $A_s$ bears the polarization information of the incident beam, $\mathbf{e}_p$ and $\mathbf{e}_s$ are the unit vectors along $p$ and $s$ polarizations, respectively.    
\begin{figure}[ht!]
\centering\includegraphics[width=0.7 \linewidth]{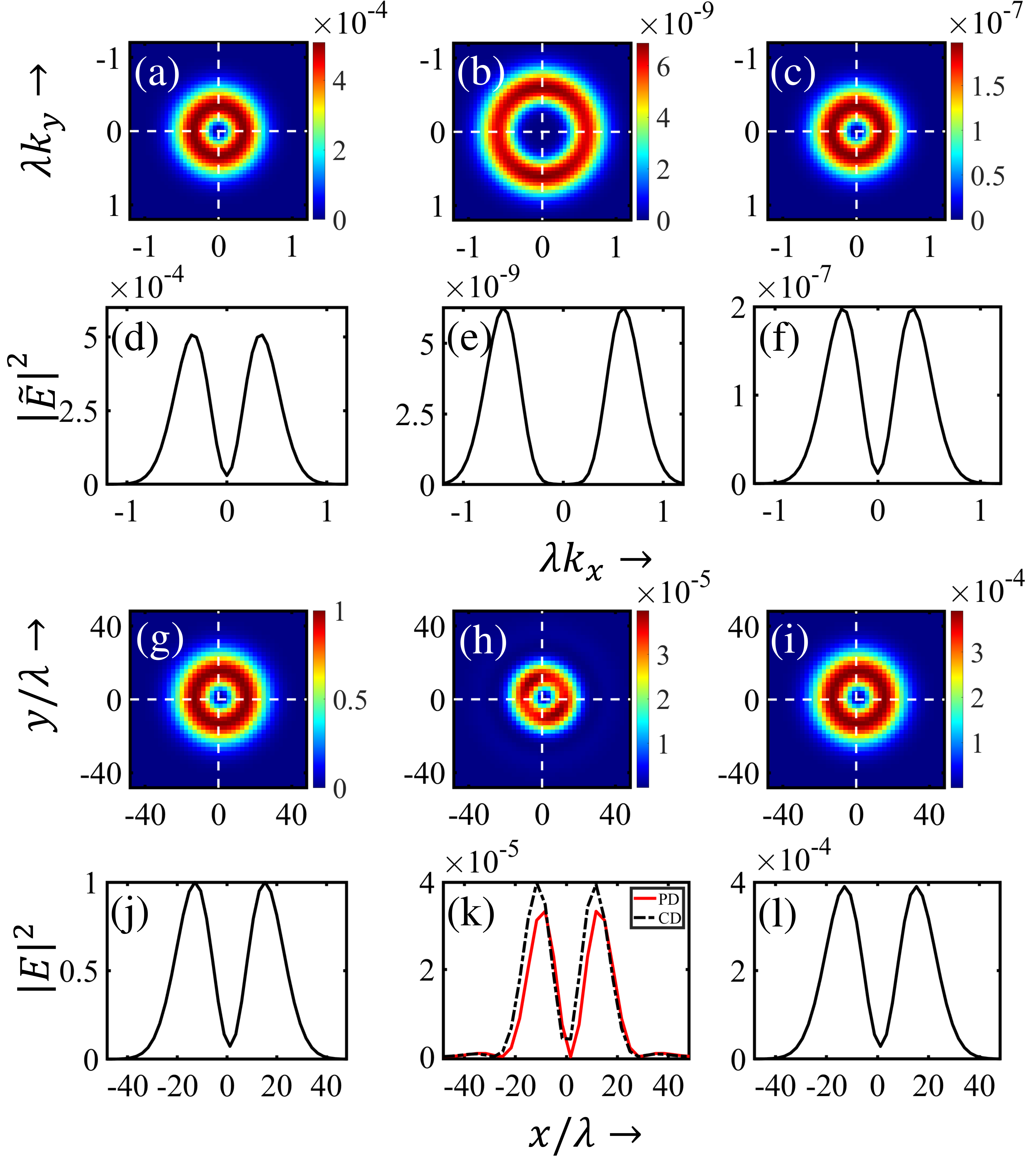}
\caption{(a) Incident, (b) reflected, and (c) transmitted spectra of the $s-$polarized LG beam incident normally on the structure. The line plots in (d)-(f) along the horizontal axis further illustrate the characteristics of these spectra presented in (a)-(c). Real-space (g) incident, (h) reflected, and (i) transmitted beams. Line plots (j) and (l) along the $x-$axis depict the nature of the beams corresponding to (g) and (i). (k) Line plots along the principal (PD) and counter diagonal (CD) of the intensity profile in (h). Parameters mentioned in Fig. \ref{fig:gaussiannormal} are used for these simulations.}
\label{fig:slgnormal}
\end{figure}
Assuming that the central wavevector makes an angle $\vartheta_i$ with the normal to the slab, the local spherical angles $(\theta_i,\phi_i)$ for the angular decomposition of an off-axis wavevector can be found in \cite{Bliokh2013,SinhaBiswas2023,BISWAS2024130766} as:
\begin{equation}\label{thetaa}
    \theta_{i}= \tan^{-1}\left(\frac{\sqrt{k_y^2+(k_x\cos{\vartheta_i}+k_z\sin{\vartheta_i})^2}}{-k_x\sin{\vartheta_i}+k_z\cos{\vartheta_i}}\right)
\end{equation}
\begin{equation}\label{phii}
	\phi_i = \tan^{-1}\left(\frac{k_y}{k_x\cos{\vartheta_i}+k_z\sin{\vartheta_i}}\right)
\end{equation}
Where, $k_z=\sqrt{k_0^2-(k_x^2+k_y^2)}$. Eqs. \ref{thetaa}  and \ref{phii} also yield the polar and  azimuthal angles of the reflected and transmitted wavevectors; however,  $\vartheta_i$ should be replaced with the angle of reflection $\vartheta_r$ and the angle of refraction $\vartheta_t$ of the central $k$-vector, respectively. Rotational transformations from the beam frame to the $p$ and $s$ polarization modes, for each of the spatial harmonics (or $k$-vectors), are performed with the corresponding polar and azimuth angles which are subsequently associated with the reflection ($r_p$ and $r_s$) and transmission ($t_p$ and $t_s$) coefficients via transfer matrixa to get the scattered amplitudes $|\mathbf{E}_a\rangle$:    
\begin{equation}\label{inverse_fourier}
    | \mathbf{E}_a \rangle = U_a^\dag F_a U_i | \mathbf{E}_i \rangle
\end{equation}
Where,
\begin{equation}\label{rotation}
    U_a = \hat{R}_y(\theta_a) \hat{R}_z(\phi_a) \hat{R}_y(\vartheta_a)
\end{equation}
\begin{equation}
    F_a = diag(a_p,a_s),     a = r,t   
\end{equation}
\begin{figure}[ht!]
\centering\includegraphics[width=0.7 \linewidth]{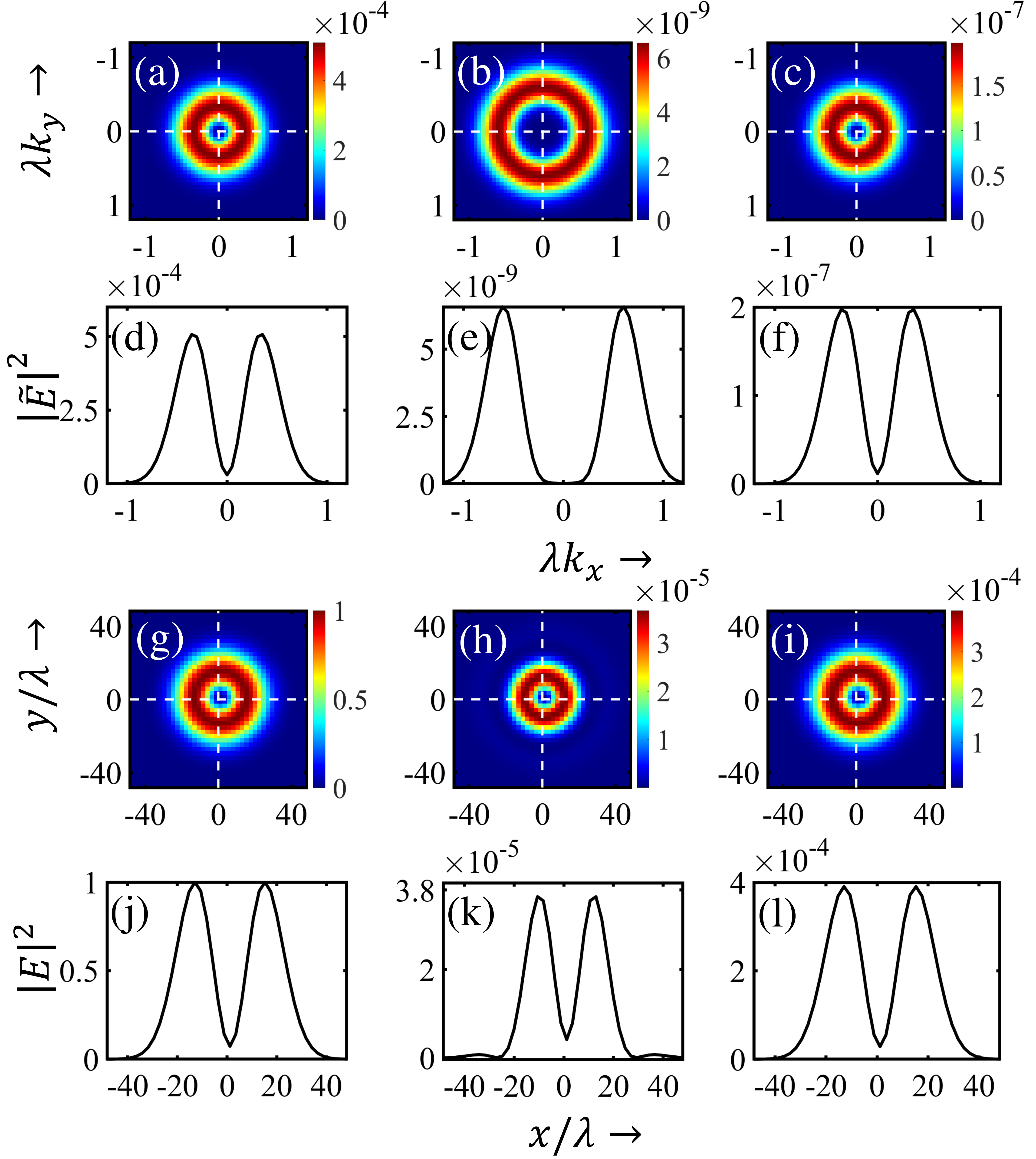}
\caption{(a) Incident, (b) reflected, and (c) transmitted spectra of the left circularly polarized LG beam incident normally on the structure. The line plots in (d)-(f) along the horizontal axis illustrate the characteristics of these spectra presented in (a)-(c). Real-space (g) incident, (h) reflected, and (i) transmitted beams. Line plots (j)-(l) along the $x-$axis depict the nature of the beams corresponding to (g)-(i). Parameters mentioned in Fig. \ref{fig:gaussiannormal} are used for these simulations.}
\label{fig:lcplgnormal}
\end{figure}
\begin{figure}[ht!]
\centering\includegraphics[width=0.7 \linewidth]{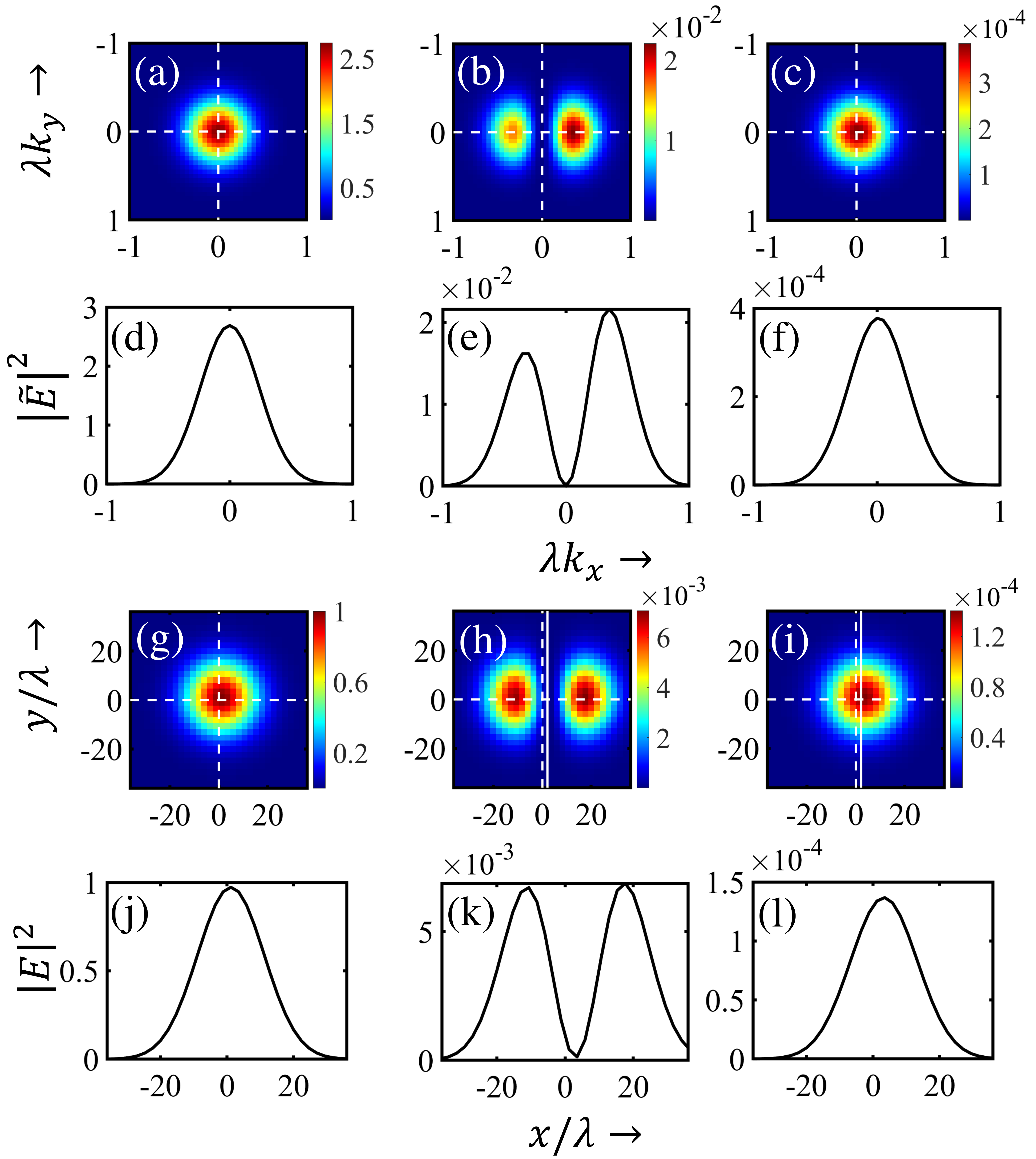}
\caption{(a) Incident, (b) reflected, and (c) transmitted spectra of the $s-$polarized Gaussian beam incident obliquely at an angle of $45^{\circ}$. The line plots in (d)-(f) along the horizontal axis further illustrate the characteristics of these spectra presented in (a)-(c). Real-space (g) incident, (h) reflected, and (i) transmitted beams. Line plots (j)-(l) along the $x-$axis depict the nature of the beams corresponding to (g)-(i). White solid lines in (h) and (i) denote the GH (along $x$) shifts of the field profiles. Parameters are: $\epsilon_i=1.0$, $\epsilon_2=2.6244$, $\epsilon_h=2.25$, $\epsilon_a=5.7121$, $\epsilon_b=2.6244$, $\epsilon_f=2.25$, $d_1=0.01\mu m$, $d_2=0.754\mu m$, $f_m=0.05$, N=10.}
\label{fig:sgaussianoblique}
\end{figure}

\begin{figure}[ht!]
\centering\includegraphics[width=0.7 \linewidth]{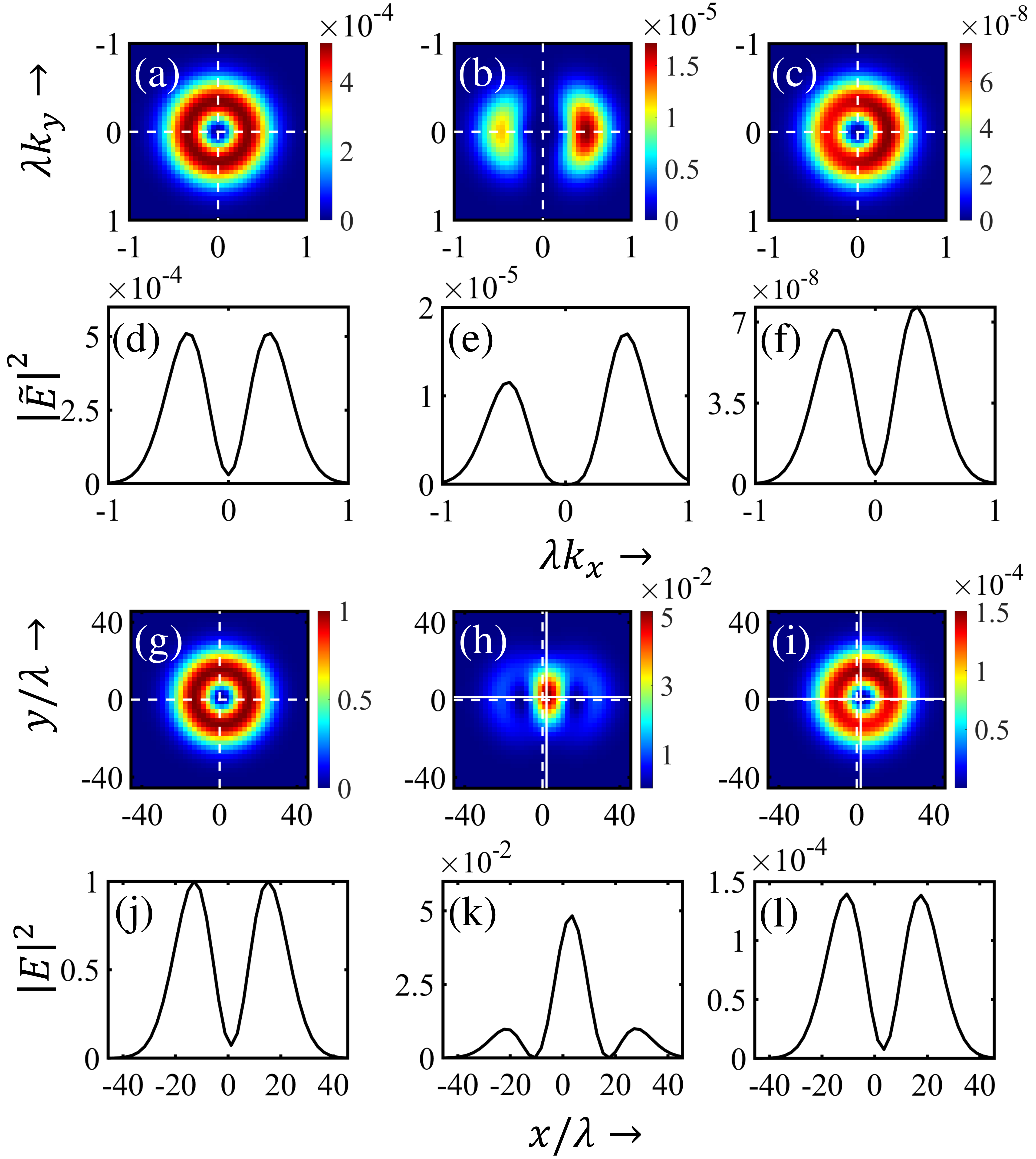}
\caption{(a) Incident, (b) reflected, and (c) transmitted spectra of the $s-$polarized LG beam incident obliquely at an angle of $45^{\circ}$. The line plots in (d)-(f) along the horizontal axis further illustrate the characteristics of these spectra presented in (a)-(c). Real-space (g) incident, (h) reflected, and (i) transmitted beams. Line plots (j)-(l) along the $x-$axis depict the nature of the beams corresponding to (g)-(i). White solid lines in (h) and (i) denote the GH (along $x$)and IF (along $y$) shifts of the field profiles. Parameters mentioned in Fig. \ref{fig:sgaussianoblique} are used for these simulations.}
\label{fig:slgoblique}
\end{figure}
The subscripts $i$, $r$, and $t$ stand for the incident, reflected, and transmitted beams respectively. $R_{x,y,z}$ in Eq. \ref{rotation} are the $3\times3$ rotation matrices along $x$, $y$, and $z$ axis respectively in the lab frame. The real space beam profiles are found by evaluating the inverse Fourier transforms of the spectrum given by Eq. \ref{inverse_fourier}. 
While examining the LG beams, the spectrum in Eq. \ref{gaussain_spectrum} needs to be replaced by the LG beam's spectrum \cite{Bliokh2013}:
\begin{equation}
    |\mathbf{E}_i\rangle = \frac{w_0}{\sqrt{2\pi}} e^{\{-(k_0w_0)^2\theta_z^2/4\}}\theta_z^{|l|}e^{il\phi+ik_0(1-\theta_z^2/2)z}(A_p\mathbf{e}_p + A_s\mathbf{e}_s)
\end{equation}
where, $\theta_z=\sqrt{k_x^2+k_y^2}/k_0$, $\phi=\tan^{-1}(k_y/k_x)$, and the azimuthal phase factor $e^{il\phi}$ contains the vortex charge $l=0,\pm1,\pm2,...$.

The in-plane ($x^{a}$) and out-of-plane ($y^{a}$) shifts are quantified as the shifts of the center of gravity of the reflected and transmitted beams. These shifts are calculated in terms of field components \cite{Bliokh2013}, as:  
\begin{equation}\label{ghshift}
    x^{a}=\frac{\int x|\mathbf{E}_{a}|^2 dx dy }{\int |\mathbf{E}_{a}|^2 dx dy }
\end{equation}
\begin{equation}\label{ifshift}
    y^{a}=\frac{\int y|\mathbf{E}_{a}|^2 dx dy }{\int |\mathbf{E}_{a}|^2 dx dy }
\end{equation}
Where, $a=r,t$ denotes the reflected and transmitted beams respectively.

\section{Results and discussions} 

\subsection{Role of the composite material}
We have plotted the real and imaginary parts of the dielectric function of the composite in Figs. \ref{response} (b), (c), respectively, for three volume fractions of silver, $f_m=0.05, 0.01, 0.1$. The dielectric function $\epsilon_m$ of silver is obtained by interpolating the experimental data from Johnson and Christy \cite{PhysRevB.6.4370}. The other parameters are taken as: N=10, $\epsilon_i=1.0$ (refractive index of air), $\epsilon_h=2.25$, and $\epsilon_f=2.25$. Note that the bulk data for silver are applicable to nanoparticles as small as $4 nm$. It is clear from Fig. \ref{response} (c) that bulk silver (line plots corresponding to $f_m=1$) does not exhibit any resonance features at around $400 nm$, while the composite shows prominent localized plasmon resonances. The other important feature is the dependence of the resonances on the volume fraction $f_m$. The increase in oscillator strength (peak absorption) accompanied by a gradual shift of the resonance towards larger wavelengths with increasing volume fraction $f_m$ of the metal (silver) shows the tunability of the absorption of the composite material. For the rest of the study, we have used volume fraction, $f_m=0.05$, and the thickness of the composite material, $d_1=0.01\mu m$, both for cases corresponding to normal and oblique incidence. Fig. \ref{response}(d) shows the critical coupling at $414nm$ for normlly incident s-polarized plane wave and Fig. \ref{response}(e) shows the same at $415nm$ for the obliquely incident plane wave, respectively. The corresponding spacer layer thicknesses are $d_2= 0.678\mu m$ (for normal incidence) and $d_2=0.754\mu m$ (for $45^\circ$  oblique incidence).
\par
In the following, we investigate the critical coupling for three different beams: 1) s-polarized Gaussian beam, 2) s-polarized LG beam with $l=\pm 1$, and 3) left circularly polarized (LCP) LG beam with $l=\pm 1$ -- for normal and $45^{\circ}$ angle of incidence. For each angle of incidence, we have searched for the optimized thicknesses of different layers for the central plane wave in the spectrum and used the same to investigate the possibility of critical coupling of the whole beam. It should be noted that for each angle of incidence, multiple thicknesses of the spacer layer can produce $R+T \approx 0$.




\subsection{Normal incidence of beams}
We first consider an s-polarized Gaussian beam with a beam waist $w_0=10\lambda$ incident on the structure normally. For the central plane wave in the spectrum, critical coupling is achieved for the spacer layer thickness $d_2=0.678 \mu m$ and at a wavelength $\lambda=414 nm$. Figure \ref{response}(d) showcases the reflectivity ($R$), transmissivity ($T$), and the total scattering ($R+T$) as a function of wavelength. The spectrum and the real space profile of the incoming Gaussian beam are shown in Figs. \ref{fig:gaussiannormal}(a), (g). Line plots along the horizontal axis (Figs. \ref{fig:gaussiannormal}(d), (j)) highlight their Gaussian nature. Due to near destructive interference of the central $k$-vectors of the reflected components from different interfaces of the multilayer, a central hollow region is observed in the spectrum of the reflected beam (Fig. \ref{fig:gaussiannormal}(b)). The line plot in Fig. \ref{fig:gaussiannormal}(e) highlights the central dip in the spectrum. Consequently, the spread of the reflected spectrum appears larger than the incident one. The real-space reflected beam becomes narrower (central portion) and develops a faint concentric intensity ring (Fig. \ref{fig:gaussiannormal}(h), (k)). The faint intensity ring is more prominent in the line plot in Fig. \ref{fig:gaussiannormal}(k). However, in the transmitted beam, no such destructive interference is observed and both the transmitted spectrum (Figs. \ref{fig:gaussiannormal}(c), (f)) and the transmitted beam (Figs. \ref{fig:gaussiannormal}(i), (l)) appear Gaussian. Interestingly, the maximum intensities of the reflected and transmitted beams are observed to be different. The reflected beam ($~10^{-5}$) is weaker by one order compared to the transmitted beam ($~10^{-4}$). It is to be noted that the appearance of the central hollow in the reflected beam is not general in this case, rather it depends on the thickness of the spacer layer. If the spacer thickness is not sufficient enough to introduce a (near) $\pi$ phase (or equivalent to $\lambda/4$ amount of path difference) between the incident and reflected components, the central hollow does not appear. This phenomenon is depicted in the line plots in Fig \ref{fig:gaussiannormal}(m) with three spacer thicknesses ($d_2=0.038\mu m,0.422\mu m,0.677\mu m$), where although the $R+T\approx 0$ is achieved for a spacer thickness of $0.038\mu m$, there is no dip in the corresponding reflected spectrum. Thus, for a very thin spacer layer, the reflected beam profile may appear different (Fig. \ref{fig:gaussiannormal}(n)). Also, with an increase in the DBR sublayer period ($N$), the transmissivity of the DBR decreases, resulting in a decrease in the maximum value of the transmitted beam. Fig. \ref{fig:gaussiannormal}(o) shows the decrease in the maximum intensity of the transmitted beam over the range of $N=10$ to $N=20$. However, the maximum value of the reflected beam remains almost the same, indicating an overall enhancement in the trapping of electromagnetic energy within the multilayer structure.
\begin{figure}[ht!]
\centering\includegraphics[width=0.7 \linewidth]{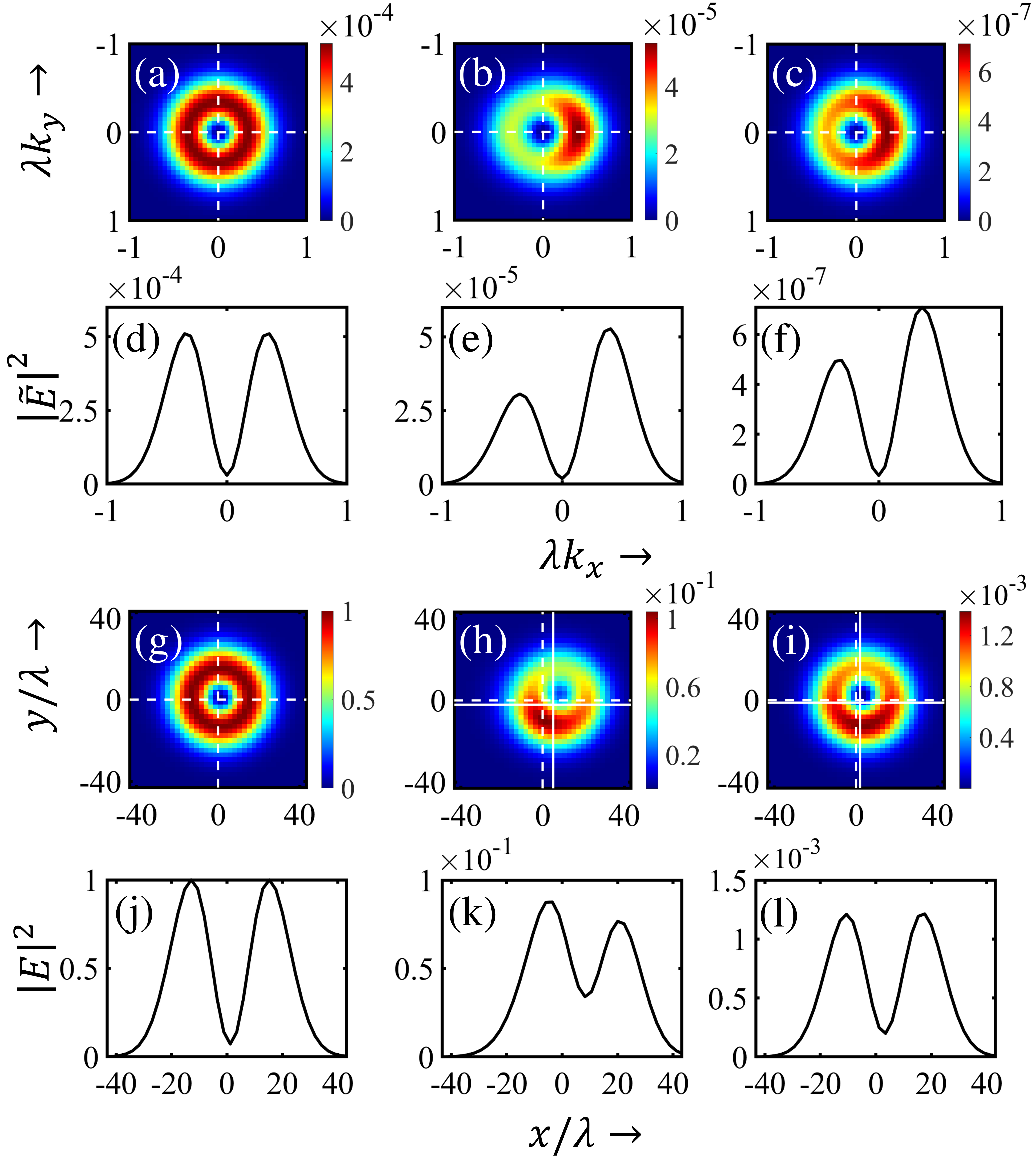}
\caption{(a) Incident, (b) reflected, and (c) transmitted spectra of the left circularly polarized LG beam incident obliquely at an angle of $45^{\circ}$. The line plots in (d)-(f) further illustrate the characteristics of these spectra presented in (a)-(c). Real-space (g) incident, (h) reflected, and (i) transmitted beams. Line plots (j)-(l) along the horizontal axis depict the nature of the beams corresponding to (g)-(i). White solid lines in (h) and (i) denote the GH (along $x$)and IF (along $y$) shifts of the field profiles. Parameters mentioned in Fig. \ref{fig:sgaussianoblique} are used for these simulations. }
\label{fig:lcplgoblique}
\end{figure}
\begin{figure}[ht!]
\centering\includegraphics[width=0.7 \linewidth]{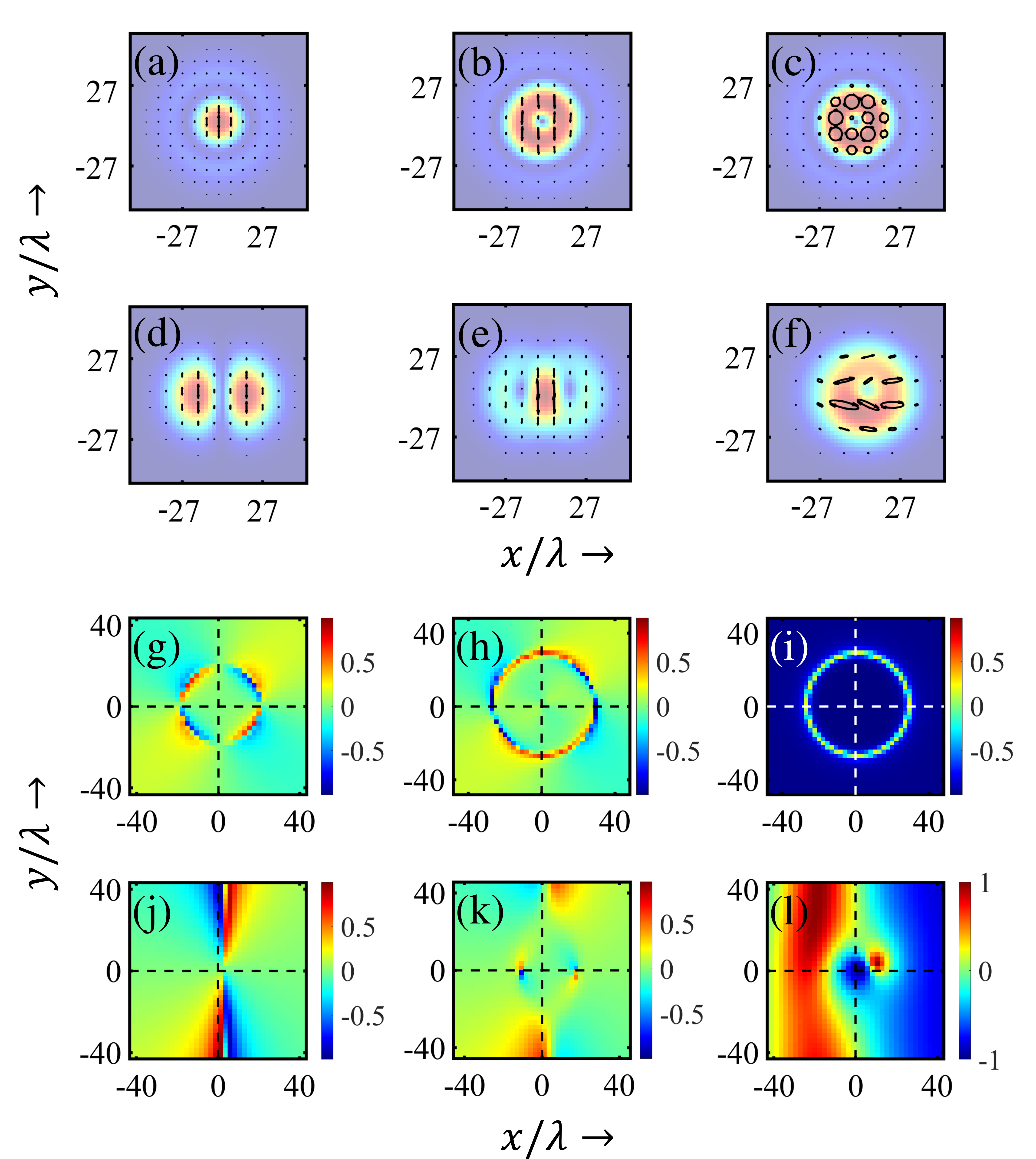}
\caption{Polarization states across the cross-section of the reflected beam for the (a) s-polarized Gaussian beam, (b) s-polarized LG beam, and (c) left circularly polarized LG beam incident normally on the structure. (d)–(f) represent the polarization states in the reflected beams under oblique incidence of the beams (in the same sequence) as mentioned earlier. The backgrounds in (a)–(f) contain the contrast-enhanced and semi-transparent intensity distributions of the reflected beams. The degree of circular polarization (normalized), $S3(=2Im(E_x^{*}E_y))$, for the normal case, across the cross-section of the reflected beam is shown for the incident (g) s-polarized Gaussian beam, (h) s-polarized LG beam, and (i) left circularly polarized LG beam, respectively. (j)–(l) represent the corresponding plots for the degrees of circular polarization for the same incident beams under oblique incidence}
\label{fig:polarimetry}
\end{figure}
Next, we consider two LG beams, one with s-polarization and the other with left circular polarization (LCP). Both the beams have orbital angular momentum (OAM) $l=+1$. The incident, reflected and transmitted beams and the corresponding spectra for the s-polarized beam are showcased in Fig. \ref{fig:slgnormal}, while those of the LCP beam are showcased in Fig. \ref{fig:lcplgnormal}. Thus, the incident beam spectra is displayed in Fig. \ref{fig:slgnormal}(a) and \ref{fig:lcplgnormal}(a), reflected spectra in Fig. \ref{fig:slgnormal}(b) and \ref{fig:lcplgnormal}(b) and transmitted spectra in Fig. \ref{fig:slgnormal}(c) and \ref{fig:lcplgnormal}(c). Similarly, incident beams are shown in Fig. \ref{fig:slgnormal}(g) and \ref{fig:lcplgnormal}(g), reflected beams in Fig. \ref{fig:slgnormal}(h) and \ref{fig:lcplgnormal}(h) and transmitted beams in Fig. \ref{fig:slgnormal}(i) and \ref{fig:lcplgnormal}(i). Corresponding line plots along the horizontal axes are shown just below the respective field distributions in Figs. \ref{fig:slgnormal}((d)-(f), (j)-(l)) and Figs. \ref{fig:lcplgnormal}((d)-(f), (j)-(l)). LG beams inherently possess a null intensity at the beam center. Because of the critical coupling in the structure, the central $k$-vectors in the spectrum destructively interfere resulting in a wider spectrum of the reflected beams as shown in Figs. \ref{fig:slgnormal}(b), (e) and Figs. \ref{fig:lcplgnormal}(b), (e). It is to be noted that for the s-polarized beam, the ring in the reflected spectrum is not uniform (Fig. \ref{fig:slgnormal}(b)). On the other hand, for a circularly polarized beam, the ring is uniform (Fig. \ref{fig:lcplgnormal}(b)). Thus, although in both cases, the reflected beam is accompanied by a faint ring because of the broadening of the spectra (Fig. \ref{fig:slgnormal}(h) and Fig. \ref{fig:lcplgnormal}(h)), the reflected beam for the s-polarization is non-uniform  (Fig. \ref{fig:slgnormal}(h)). This non-uniformity is highlighted in the line plots in Fig. \ref{fig:slgnormal}(k), where the intensities along the principal diagonal and the counter diagonal are seen to be different. The observed non-uniformity arises from azimuthal differential attenuation $(\mathscr{D})$ of orthogonal polarizations, a characteristic effect typically observed in tight focusing \cite{basudev2013,Roy2014} and scattering \cite{nirmalyspolarimetry}. Diattenuation, defined as the differential attenuation of orthogonal polarizations, is of geometric origin and hence intimately connected to the polarization and the spectrum of the light (or its spectral spread, characteristic of the beam) under investigation \cite{nirmalyspolarimetry}. Upon interaction with the multilayer structure, portions of the s(p)-polarization component are converted into p(s) polarization, resulting in scattered fields (i.e, reflected and transmitted beams) with mixed polarization. For linearly polarized (s-polarized) beams, this azimuthal diattenuation (i.e., diattenuation along the azimuthal direction) manifests as a nonuniform intensity ring in the scattered beam’s cross-section. For circularly polarized beams, $\mathscr{D}$ being zero ($\mathscr{D}=0$), no such non-uniformity appears in the scattered beams.  In the following section, we will examine its influence on the degree of circular polarization ($S3$). However, as noted earlier, the semi-analytical approach employed here does not yield closed-form field expressions, thereby restricting further analytical exploration of diattenuation. Additionally, in all the cases of normal incidence, whether with Gaussian or LG beams, the peak intensity of the reflected beam consistently maintains the same order. The ratio of the peak incident intensity to the peak reflected intensity is approximately $1:10^{-5}$. Similarly, for all beams, the peak transmitted intensity is around $10^{-4}$. It is also noteworthy that due to the rejection band of the DBR, in all cases (Fig \ref{fig:gaussiannormal}(i), \ref{fig:slgnormal}(i), \ref{fig:lcplgnormal}(i)), the nature of the spectra remains unchanged. Ideally, a single plane wave should experience no transmission. However, the intrinsic angular dependence ($\theta, \phi$) of spatial frequencies within the beam spectrum necessitates varying DBR sublayer thicknesses for different off-axis $k$-vectors to achieve complete reflectio -- a requirement that cannot be met with fixed-parameter DBR layers. Thus, the field distributions of the transmitted beams follow that of the incident beam -- Gaussian for the incident Gaussian beam and LG for the incident LG beam.

\subsection{Oblique incidence of polarized beams}

The oblique incidence of polarized beams is of immense interest, as the reflected and transmitted beams undergo (incident) angle-dependent modifications and experience in-plane (Eq. \ref{ghshift}) and out-of-plane (Eq. \ref{ifshift}) shifts. In this case, for the central plane wave, critical coupling is achieved for a spacer layer thickness of $d_2 = 0.754\mu m$ and at a wavelength $\lambda = 415nm$ (Fig. \ref{response}(e)). When the s-polarized Gaussian beam (Figs. \ref{fig:sgaussianoblique}(a), (d), (g), (j)), fall obliquely at an angle of $45^{\circ}$, the reflected spectrum is no longer Gaussian, but consists of multiple (two) distinct lobes with uneven strengths (Figs. \ref{fig:sgaussianoblique}(b), (e)). The reason behind the complex reflected spectrum lies primarily in two factors: 1) multiple reflections and refractions within the multilayer, and 2) the conditions of constructive and destructive interferences of the reflected and refracted components are now critically dependent on the angle of incidence of the beam. These two factors essentially determine the field distribution for oblique incidence of any kind of beams as will be evident in the case of polarized LG beams as well. Consequently, the reflected beam (Figs. \ref{fig:sgaussianoblique}(h), (k)) is observed to have two lobes with almost equal intensities. It is to be noted that the equal intensities in the lobes are not a general property of the reflected beam, rather, the intensities change for different angles of incidence of the beam. In contrast, both the transmitted spectrum (Figs. \ref{fig:sgaussianoblique}(c), (f)) and the transmitted beam (Figs. \ref{fig:sgaussianoblique}(i), (l)) are observed to retain the usual Gaussian profile. The in-plane GH shifts for the reflected (Fig. \ref{fig:sgaussianoblique}(h)) and transmitted (Fig. \ref{fig:sgaussianoblique}(i)) beam are found to be $x^r=2.09\lambda$ and $x^t=2.10\lambda$. No out-of-plane (or IF) shift is experienced by the reflected or the transmitted beams (see Table \ref{tab:shifts}) as expected for linearly polarized light. 

\begin{table}[ht]
    \centering
    \resizebox{0.9\textwidth}{!}{
    \begin{tabular}{|c|c|c|c|c|c|}
        \hline
        & & \multicolumn{2}{c|}{In-plane (GH) shifts} & \multicolumn{2}{c|}{Out-of-plane (or transverse) shifts} \\ \hline
        & & reflection($x^r_{\pm l}$) & transmission($x^t_{\pm l}$) & reflection($y^r_{\pm l}$) & transmission($y^t_{\pm l}$) \\ \hline
        s-pol (Gaussian) & &  2.09 &  2.10 & 0 & 0 \\ \hline
        \multirow{3}{*}{s-pol (LG)} & \(l=+1\) &2.08 & 2.11 & -1.50 & -0.47 \\ \cline{2-6}
        & \(l=-1\) &2.08 & 2.11 & 1.50 & 0.47 \\ \hline
        \multirow{2}{*}{LCP (LG)} & \(l=+1\) & 5.21 & 1.96 & -2.09 & -1.27 \\ \cline{2-6}
        & \(l=-1\) & 5.21 & 1.96 & 1.59 & 1.19 \\ \hline
        \multirow{2}{*}{LCP (LG)} & \(l=+4\) & 4.48 & 1.99 & -6.76 & -4.88 \\ \cline{2-6}
        & \(l=-4\) & 4.49 & 1.99 & 6.39 & 4.80 \\ \hline
    \end{tabular}
    }
    \caption{In-plane (GH) and out-of-plane shifts experienced by the reflected and transmitted beams for the incident s- and left circularly polarized Gaussian and LG beams. All shifts are in units of the wavelength $\lambda$ $ (415 nm) $.}
    \label{tab:shifts}
\end{table}
For an s-polarized LG beam (Figs. \ref{fig:slgoblique}(a), (d), (g), (j)) with $l=+1$ incident obliquely ($45^{\circ}$), the reflected spectrum (Figs. \ref{fig:slgoblique}(b), (e)) consists of two lobes with different strengths. The resulting reflected beam is observed to experience heavy distortion as shown in Figs. \ref{fig:slgoblique}(h), (k). Complex constructive and destructive interference among the multiple reflected and refracted components is again responsible for this heavy distortion. The transmitted beam and the corresponding spectrum are also shown in Figs. \ref{fig:slgoblique}(c), (f), (i), (l). In this case, both the reflected and the transmitted beams experience in-plane and out-of-plane shifts. The shifts for the reflected beam are given by $x^r_{+1}=2.08\lambda$, $y^r_{+1}=-1.50\lambda$ (Fig. \ref{fig:slgoblique}(h)), while for the transmitted beam, the shifts are $x^t_{+1}=2.11\lambda$,  $y^t_{+1}=-0.47\lambda$ (Fig. \ref{fig:slgoblique}(i)). Thus, the reflected and transmitted beams experience different amounts of shifts (see Table \ref{tab:shifts}).

Here, we draw attention to a distinct feature of beam shifts for beams carrying intrinsic orbital angular momentum (OAM). Under oblique incidence, the intrinsic OAM of opposite topological charges ($l$) induces vortex-dependent (or $l$-dependent) transverse shifts — a phenomenon known as the orbital Hall effect of light \cite{Bliokh2013}. As a result, beams with opposite charges shift in opposite transverse directions (see Table \ref{tab:shifts}). This becomes evident when comparing the shifts of LG beams with $l = +1$ and $l = -1$. For an LG beam with $l = -1$, the GH shifts are identical to those for $l = +1$, specifically, $x^r_{-1} = x^r_{+1} =2.08\lambda$ for the reflected beam and $x^t_{-1} = x^t_{+1} = 2.11\lambda$ for the transmitted beam. However, the transverse shifts are exactly opposite: $y^r_{-1} = -y^r_{+1} = 1.50\lambda$ for the reflected beam, and $y^t_{-1} = -y^t_{+1} = 0.47\lambda$ for the transmitted beam. It is important to note that the shifts are calculated as deviations of the beam centroid and represent the first moments of the intensity distributions. Thus, these shifts remain a valid measure for characterizing the reflected and transmitted beams, even in the presence of distortions.


For a left circularly polarized LG beam with $l=+1$ falling obliquely at an angle of $45^{\circ}$, its spectrum and the real space beam profiles are shown in Figs. \ref{fig:lcplgoblique}(a), (d), (g), (j). Both the reflected (Figs. \ref{fig:lcplgoblique}(b), (e)) and transmitted (Figs. \ref{fig:lcplgoblique}(c), (f)) spectra undergo significant modifications which in turn yield distorted reflected (Figs. \ref{fig:lcplgoblique}(h), (k)) and transmitted beams (Figs. \ref{fig:lcplgoblique}(i), (l)). It is to be noted that both the transmitted and reflected beams experience in-plane (GH) and out-of-plane shifts, (see Table \ref{tab:shifts}) with the net resulting shifts being $x^r_{+1}=5.21\lambda$ and $y^r_{+1}=-2.09\lambda$ for the reflected beam and $x^t_{+1}=1.96\lambda$ and $y^t_{+1}=-1.27\lambda$ for the transmitted beam. Here, another aspect of beam shift appears, resulting from the combined effect of the circular polarization and the intrinsic OAM (or vortex charge, $l$) of the beams. In general, for linearly polarized beams carrying orbital angular momentum (OAM), the vortex-induced transverse shifts are exactly opposite for opposite OAM. However, for beams with the same circular polarization but with opposite OAM, the transverse ($y$) shifts differ for opposite values of $l$. To elaborate on the phenomenon, we simulate the reflected and transmitted beams for a left circularly polarized LG beam with $l=-1$, i.e., the beam with the same helicity as considered above but with a negative value of $l$. In this case, the net beam shifts in the reflected beam are found to be $x^r_{-1}=5.21\lambda$ and $y^r_{-1}=1.59\lambda$, while those in the transmitted beam are $x^t_{-1}=1.96\lambda$ and $y^t_{-1}=1.19\lambda$. Thus, beams with the same helicity but with opposite intrinsic OAM, experience different amounts of transverse shifts, $|y^r_{+1}| \neq |y^r_{-1}|$ and $|y^t_{+1}| \neq |y^t_{-1}|$ (see Table \ref{tab:shifts}). The underlying reason for different shifts is that the vortex-induced shifts can either increase or decrease the shifts due to the polarization of the beam. Notably, the lack of closed-form expressions for the electric fields hinders the separate characterization of total beam shifts in terms of their polarization and vortex contributions. The effect is more pronounced for beams with higher orbital angular momentum. Here, we detail the shifts for an LG beam with $l=\pm4$ (see Table \ref{tab:shifts}). For the beam with $l=+4$, the shifts are $x^r_{+4}=4.48\lambda, y^r_{+4}=-6.76\lambda$ for the reflected beam and $x^t_{+4}=1.99\lambda, y^t_{+4}=-4.88\lambda$ for the transmitted beam. Similarly, for the beam with $l=-4$, the shifts are $x^r_{-4}=4.49\lambda, y^r_{-4}=6.39\lambda$ for the reflected beam and $x^t_{-4}=1.99\lambda, y^t_{-4}=4.80\lambda$ for the transmitted beam.



So, for the oblique incidence case, both the Gaussian and LG beam experience significant distortions. The distortions are more prominent in the reflected beam than in the transmitted beam. The combined effect of circular polarization and the vorticity of an LG beam makes the beams shift complicated. Such oblique incidence problems in the context of mirrors using DBR have also been investigated in a very recent work in Ref.~\cite{beamshiftdbr}. Overall, due to the multiple reflections and refractions combined with polarization- and vortex-dependent shifts experienced by the beams, critical coupling for polarized beams in this multilayer system cannot be achieved perfectly.




\subsection{Degree of circular polarization}

The state of polarization is an important quantity for characterizing reflected and transmitted beams, particularly in the context of the polarization-dependent interaction of incident vector beams with a multilayer structure. As mentioned earlier, due to the interaction with the structure, partial inter-conversions between the s- and p-polarizations disrupt the uniformity of the state of polarization across the beam cross-sections. As a result, the reflected or transmitted beams possess mixed polarizations. For normal incidence, the states of polarization of the reflected beams predominantly follow the states of polarization of the incident beam: linear for linearly polarized beams (s-polarized Gaussian beam in Fig. \ref{fig:polarimetry}(a) and s-polarized LG beam in Fig. \ref{fig:polarimetry}(b)) and circular for circularly polarized beams (Fig. \ref{fig:polarimetry}(c)). However, for oblique incidence, the depolarization effects are more significant, as seen in Fig. \ref{fig:polarimetry}(e) for the s-polarized LG beam and in Fig. \ref{fig:polarimetry}(f) for the left circularly polarized LG beam. For the s-polarized Gaussian beam, the polarization remains linearly polarized even for oblique incidence (Fig. \ref{fig:polarimetry}(d)).

Additionally, polarization mixing gives rise to circular polarization in the reflected or transmitted beams. For beams normally incident, the normalized degrees of circular polarization ($S3$) for the reflected beams are shown in Figs. \ref{fig:polarimetry}(g), (h), (i) for s-polarized Gaussian beams, s-polarized LG beams, and LCP LG beams, respectively. The corresponding distributions under oblique incidence for the respective beams are shown in Figs. \ref{fig:polarimetry}(j), (k), (l). In all the plots, the positive and negative regions of circular polarizations are juxtaposed. Azimuthal diattenuation governs the formation of circular polarization regions across the cross-section of the beam for both normal and oblique incidence, with the latter exhibiting greater complexity. In summary, the amount of polarization mixing, ellipticity, and degree of circular polarization depends on the incident polarization, the OAM content of the beam ($l$-value), and the angle of incidence of the beam. However, as mentioned earlier, the lack of an analytical form for the field expressions hinders further investigations of these quantities.
\begin{figure}[ht!]
\centering\includegraphics[width=0.85 \linewidth]{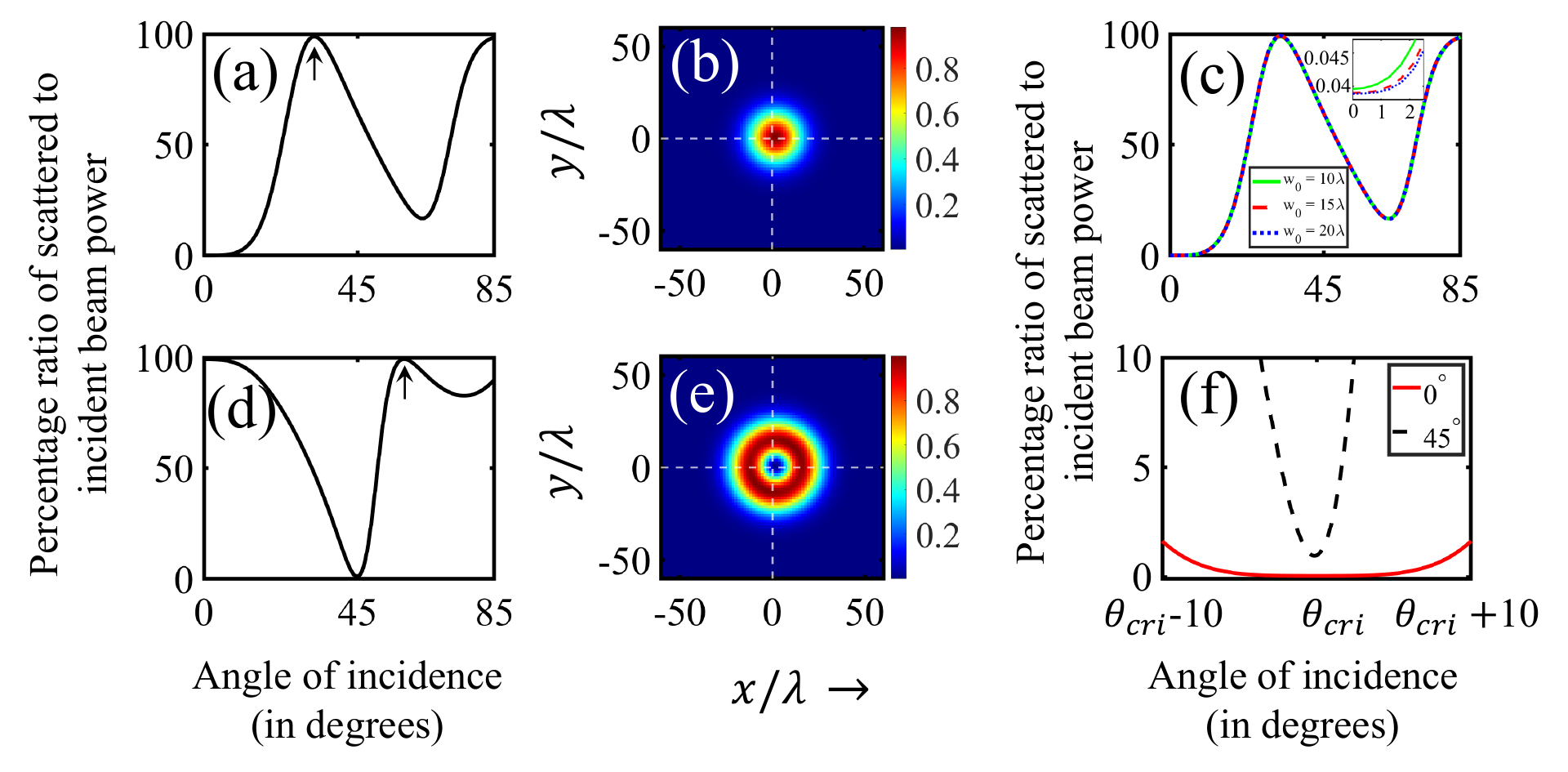}
\caption{Panels (a–c) show results for normal incidence, while panels (d–f) correspond to oblique incidence at $45^{\circ}$, (a) Angular variation of the scattered-to-incident power ratio (in \%) for an s-polarized Gaussian beam at normal incidence. (b) Real space reflected beam profile at the super-scattering angle $\theta=32.6^{\circ}$ (denoted by `$\uparrow$' in (a)), (c) Power ratio for three normally incident Gaussian beams with different beam waists: $w_0=10\lambda, 15\lambda, 20\lambda$. Inset: Magnified view of the same curves for low angle of incidences. (d) Angular variation of the scattered-to-incident power ratio (in \%) for an s-polarized Gaussian beam at oblique incidence ($45^{\circ}$). (e) Reflected real-space profile of an LG beam with $l=1$ at the super-scattering angle $\theta=58.4^{\circ}$ (denoted by `$\uparrow$' in (d)). (f) Line plot comparing the power ratio for the normally ($\theta_{cri}=0^{\circ}$) and obliquely ($\theta_{cri}=45^{\circ}$) incident s-polarized Gaussian beam.}
\label{fig:power}
\end{figure}

\subsection{Critical coupling efficiency of the structure:}
Throughout the study, critical coupling is assessed by the plane wave condition $R+T=0$. Since this condition holds only for the central plane wave, we extend the analysis to the total power scattered from the multilayer absorber in order to assess the coupling efficiency for structured beams. The powers of the incident, reflected, and transmitted beams are calculated separately using Parseval’s theorem: $P=\int\int \tilde{E}^{*}\tilde{E} \,dk_{x} \,dk_{y}$. The scattered power is then obtained as the sum of the reflected and transmitted powers. Since we are interested in the scattered-to-incident power ratio, any normalization factor in Parseval's theorem is automatically accounted for. For a normally incident s-polarized Gaussian beam, Fig. \ref{fig:power} (a) shows the variation of the scattered-to-incident power ratio (in \%) with angle of incidence, using the same structure that achieves critical coupling for a single plane wave at normal incidence. The ratio remains close to zero at normal incidence, indicating efficient coupling, and retains this behavior up to about $10^{\circ}$. Beyond this, super-scattering appears at $\vartheta_i=32.6^{\circ}$, arising from constructive interference of the reflected beams from various interfaces, all being in the same phase. This indicates that, above $10^{\circ}$, the structure no longer supports efficient coupling and instead can reflect the incoming beam with nearly $100\%$ efficiency near the super scattering angle. The corresponding real-space field profile is shown in Fig. \ref{fig:power} (b). Grazing incidence and near-grazing angles have been deliberately excluded from consideration. Other polarizations exhibit the same behavior as a function of the incidence angle. 

The dependence of coupling efficiency on beam waist is also examined. While the variation of the power ratio with angle is qualitatively similar for different waists, larger beams (being closer to plane waves) exhibit slightly higher absorption, as expected. Fig. \ref{fig:power} (c) presents the angular dependence of the scattered-to-incident power ratio for Gaussian beams with waists $w_0=10\lambda, 15\lambda$, and $20\lambda$. The inset highlights that the absorption varies monotonically with beam waist, with the widest beam ($w_0=20\lambda$) achieving the strongest absorption, followed by $w_0=15\lambda$, with $w_0=10\lambda$ absorbing the least.

For oblique incidence ($45^{\circ}$) of an $s$-polarized Gaussian beam, Fig. \ref{fig:power} (a) illustrates the variation of the scattered-to-incident power ratio (in \%) with the angle of incidence, using a structure designed to achieve critical coupling for a single plane wave at $\vartheta_i=45^{\circ}$. A pronounced near-zero power ratio is observed at $\vartheta_i=45^{\circ}$, while a distinct super-scattering phenomenon emerges at $\vartheta_i=58.4^{\circ}$. This demonstrates that a fixed planar absorbing structure can exhibit both critical coupling, corresponding to nearly zero scattering, and super-scattering, corresponding to maximal scattering. Consequently, the coupling efficiency generally spans the entire range from $\sim 0\%$ to $\sim 100\%$. The real-space reflected beam profile of an incident LG beam at the super-scattering angle (i.e., $\vartheta_i= 58.4^{\circ}$) is shown in Fig. \ref{fig:power} (e).  We also investigated the effect of vortex charge on the power ratio (not shown). Critical coupling is found to be the most efficient for $l=1$, with the efficiency decreasing with increasing $l$ value.  

For a comprehensive comparison between the two structures -- one optimized for normal incidence ($\vartheta_i=\theta_{cri}=0^{\circ}$) and the other for oblique incidence ($\vartheta_i=\theta_{cri}=45^{\circ}$) -- the scattered-to-incident beam power ratios are plotted together in Fig. \ref{fig:power} (f). The plot reveals two key observations: (1) the ratio remaining closer to zero signifies superior coupling efficiency for the structure optimized at normal incidence, and (2) the angular range over which this ratio stays near zero is again broader for the normal-incidence-optimized structure.

\section{Conclusion} 
In conclusion, this study revisits the phenomenon of critical coupling in realistic scenarios involving various incident polarized beams. The beams considered here are an s-polarized Gaussian beam, an s-polarized LG beam with $l = \pm 1$, and a left circularly polarized (LCP) LG beam with $l = \pm 1$. However, the study can be easily extended to various other structured beams. Through the rigorous vector angular spectrum method, we examined how the beam polarization and the intrinsic orbital angular momentum (OAM) influence interactions with a multilayer structure under both normal and oblique incidence. Our results indicate that, for this configuration, complete beam absorption remains unattainable, revealing an inherent limitation of the critical coupling mechanism for planar multilayer structures. Additionally, the remnant beams leaking out of the structure suffer heavy distortions and possess a mixed state of polarization across the beam cross-section. For LG beams, the interplay of circular polarization and the intrinsic orbital angular momentum leads to complex transverse beam shifts. Interestingly, we also report super-scattering from the same structures when one has constructive interference of the various reflected components. Most importantly, our treatment provides a complete and realistic approach to the problem of critical coupling to improve experimental designs in real applications and implementations. In addition, our approach also provides a foundation for further exploration, including extensions to photonic crystals, metasurfaces, nonlinear media, and the generation of structured beams for low-power applications. We anticipate that our findings will drive future research on critical coupling in emerging platforms, including graphene and other 2D materials, using fully vectorial beams.

\section{Data availability statement}
All data that support the findings of this study are included within the article.

\section{Acknowledgment}
Sauvik Roy is thankful to the Department of Science and Technology (DST), Government of India, for the INSPIRE fellowship.

\section{References}
\providecommand{\noopsort}[1]{}\providecommand{\singleletter}[1]{#1}%


\end{document}